\documentclass[final]{ias2}

\usepackage{graphicx} 
\usepackage{multirow}
\usepackage{array} 
\usepackage{hyperref} 
\usepackage{xcolor}
\usepackage{lineno}
\begin{document}
\markboth{Simulation Studies for the Tin Bolometer Array}{V Singh, et al.}

\title{Simulation studies for the Tin Bolometer Array for Neutrinoless Double Beta Decay}

\author[INOTIFR,HBNI]{V Singh} 
\author[INOTIFR,HBNI]{N Dokania}
\author[INOTIFR,HBNI]{S Mathimalar} 
\author[DNAPTIFR]{V Nanal}
\email{nanal@tifr.res.in}
\author[DNAPTIFR]{R G Pillay}
\address[INOTIFR]{India-based Neutrino Observatory, Tata Institute of Fundamental Research, Mumbai - 400005, India}
\address[HBNI]{Homi Bhabha National Institute, Anushaktinagar, Mumbai - 400094, India}
\address[DNAPTIFR]{Department of Nuclear and Atomic Physics, Tata Institute of Fundamental Research, Mumbai - 400005, India}

\begin{abstract}
It is important to identify and reduce the gamma radiation which can be a significant source of background for any double beta decay experiment. The TIN.TIN detector array, which is under development for the search of Neutrinoless Double Beta Decay in $^{124}$Sn, has the potential to utilize the hit multiplicity information to discriminate the gamma background from the events of interest. Monte Carlo simulations for optimizing the design of a Tin detector module has been performed by varying element sizes with an emphasis on the gamma background reduction capabilities of the detector array. 
\end{abstract}


 
\maketitle

\section{Introduction}

The observation of neutrino oscillations~\cite{SuperK_Atmospheric_sy, 
SNO_Solar_sy,KamLAND_Reactor_sy,K2K_accelerator_sy, MINOS_accelerator_sy}, which 
implies finite neutrino mass, is one of the most important discoveries 
in particle physics in recent years. However, the absolute mass of 
neutrinos and its true nature are not known yet. Neutrinos can either 
be Dirac (with particles and antiparticles being distinct) or Majorana 
(with particles and antiparticles being indistinguishable) particles. 
Understanding the nature of neutrinos is of fundamental importance to 
explain the origin of small neutrino masses and possibly to elucidate 
the matter-antimatter asymmetry observed in nature~\cite{LeptoGenesis_sy}.
Neutrinoless double beta  ($0\nu\beta\beta$) decay is perhaps the only feasible experiment which can probe the true nature of neutrinos and {is being pursued vigorously worldwide}~\cite{RMPReview_2008,RPPTheory_2012,AHEPReview_2014}. A feasibility study to search for $0\nu\beta\beta$ decay in $^{124}$Sn has been initiated in India. The TIN.TIN (\underline{T}he \underline{IN}dia-based \underline{TIN}{) detector} will {use the cryogenic} bolometer technique to {measure} the sum energy spectrum of the two electrons emitted at the $Q$ value of the double beta decay transition ($Q_{\beta\beta}$). The experiment will be housed at {the India based Neutrino Observatory (INO)}, an upcoming underground {laboratory} with $\sim$ 1000 m rock cover all around~\cite{Naba_Pramana}.

The sensitivity of a 
$0\nu\beta\beta$ experiment critically depends on the active mass of 
the detector, the background level and the resolution of the detector. 
Since the sensitivity increases linearly with mass, increasing the mass 
of the detector is the easiest way of improving 
the sensitivity. The TIN.TIN detector will employ a modular structure wherein a 
closely packed array of detector modules will be operated at cryogenic 
temperature. Each module itself will consist of several detector elements. It should be mentioned that the size of the individual detector element is also constrained by the calorimetry requirements, number of sensors, associated wirings and readout 
electronics~\cite{Nanal_INPC13}. The granularity of the detection volume can be used for the identification of physics processes, which may help in discrimination of 
the background events. Therefore, the structure of the array and the size of the individual detector element needs to be designed based on the effectiveness of detector granularity to discriminate multi-site events from double beta decay events (which originate at a 
single site). {In this paper we report the results of Monte Carlo simulations aimed at optimizing the design of a Tin detector module by varying element sizes to reduce  the background arising from multi-hit events due to the ambient gamma rays.}

\section{Optimization of the detector element size for the background reduction}

The experimental signature of $0\nu\beta\beta$ decay consists of measuring the sum of the kinetic energies of the electrons, which is equal to the $Q$ value of the double beta decay 
($Q_{\beta\beta}$=2292.64$\pm$0.3 keV for $^{124}$Sn). The sensitivity of the detector is critically dependent on the reduction of background. While the cosmogenic background is significantly reduced in underground laboratories, the gamma and neutron background originating from surrounding rocks can be substantially reduced by suitable shielding around the detector. However, the background contribution from the decay of the radioactive trace impurities present in the detector, peripheral materials and the shield cannot be completely eliminated. Typically, $\alpha$ and $\beta$ emitting isotopes of Thorium and Uranium decay chains on or near the surface of the detector contributes to the background and can be minimized mainly by reducing the surface contamination of the detector. Equally important, is the discrimination of the $\gamma$ background from the $0\nu\beta\beta$ events of interest. Emitted electrons will dominantly deposit their energy in the Tin detector element due to their short range, while gamma-rays can give higher multiplicities. The size of the individual element should be chosen such that the detector dimensions are large compared to the range of the electrons, thereby increasing the probability to contain the $0\nu\beta\beta$ events within the element. {The typical} range of 2 MeV electrons in tin is of the order of few millimeters, and hence $0\nu\beta\beta$ events can be contained within a small volume of the detector element. Therefore, the size of the individual detector element can range from a few {$cm^3$ to hundreds of $cm^3$}. Moreover, it is desirable to have a smaller surface-to-volume ratio as it reduces the background per unit mass originating from the surface background~\cite{Pavan}. 

Gamma-rays, resulting from natural decay chain or neutron induced reactions, have energies varying from 100 keV to 5 MeV. Photons from the decays of $^{208}$Tl, $^{214}$Bi (end products of natural radioactive decay chains) etc., dominate the background in the 
region of interest. In this energy range, the gamma-rays predominantly 
interact via Compton scattering. The absorption length of these high 
energy photons in Tin is of the order of {cms}. Unlike 
electrons, photons would typically interact with more than one detector 
element and {may} deposit only a fraction of the total energy in a 
single element detector. It is therefore possible to use the hit multiplicity ($M$) to 
discriminate between electrons and gamma-rays in a 
limited manner. If the 
multiplicity of an event is denoted by M, then total photon events detected in 
the module can be written as
    
\begin{equation}
N_{total} = (N^p + N^c)_{M=1} + (N^p + N^c)_{M > 1} 
\label{multiplicity}
\end{equation}

where $N^p$ and $N^c$ are the photopeak and the Compton scattered events, 
respectively. The $M>1$ events are expected to predominantly arise 
from photons and can be rejected during analysis with the multiplicity 
condition. Photons with $M=1$ 
can be clearly identified if it is a photopeak event ($N^p_{M=1}$). Difficulty arises 
for identification and rejection of Compton scattered $M=1$ photon 
events ($N^c_{M=1}$). It is thus essential to choose an array configuration 
where the $N^c_{M=1}$ is minimized. Since the energy resolution of the 
bolometer is expected to be better than 10 keV, the background in the region of interest 
for $0\nu\beta\beta$ decay mainly arises from the Compton scattering 
of higher energy ($>Q_{\beta\beta}$) gamma-rays. It should be mentioned that the summing of low energy gamma-rays can also contribute to the background in the region of interest.
              
\begin{figure}[ht]
\begin{center}
\includegraphics[scale=0.3]{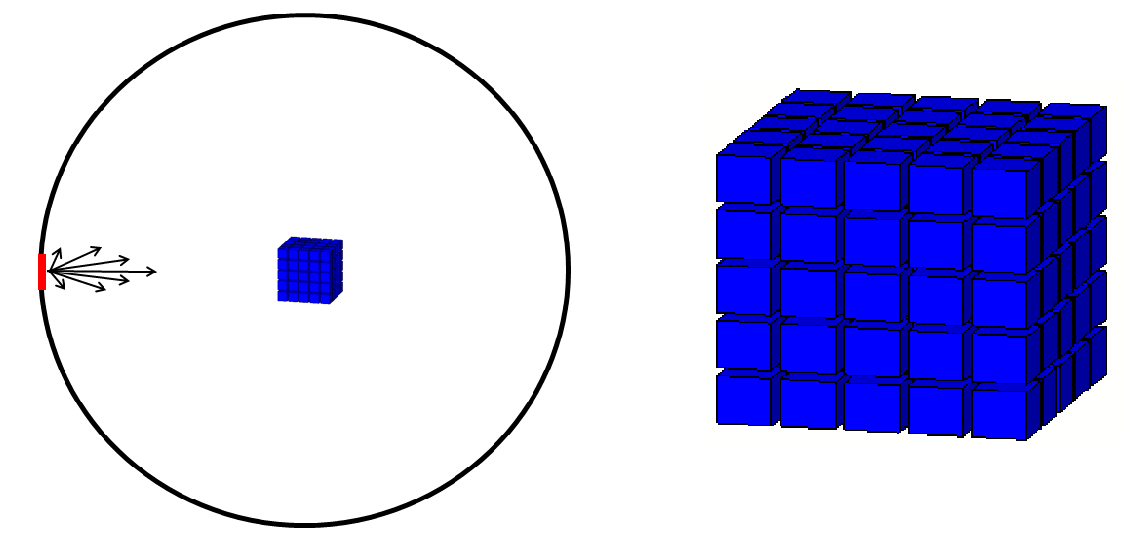}
\end{center}
\caption[Element geometry for simulation]{A pictorial view of the spherical 
surface source enclosing a 3D array of cubic Tin detector elements (left). 
A typical geometry of the element array used in simulation (right). 
The radius of the source sphere is much larger than the module size.}     
\label{simul_geometry}
\end{figure}

Simulations have been carried out to study the background resulting 
from gamma-ray 
interactions for different element configurations to optimize the size 
of the detector element and the module. The GEANT4~\cite{GEANT4} package was 
used for the simulation of particle tracking, geometries and 
physics processes. Photons of a given energy were randomly generated on a spherical 
surface  enclosing a 3D array of cubic Tin detector elements of different 
sizes. Details of the module geometry  are given in 
Table~\ref{array_simu_geometry} and shown schematically in 
Figure~\ref{simul_geometry}. 
\begin{table}[ht]
\caption[Detector module configuration used in simulations]{Detector module configuration used in simulations} 
\centering
\begin{tabular}{c c c}
\hline\hline
Individual element size & Total number of elements & Total Volume \\[0.5ex]
(cc) & & (cc) \\[0.5ex]
\hline\hline
2.143 x 2.143 x 2.143 & 7 x 7 x 7 = 343  & 3375.6 \\
3 x 3 x 3  & 5 x 5 x 5 = 125 & 3375 \\
5 x 5 x 5  & 3 x 3 x 3 = 27 & 3375 \\ 
\hline
\end{tabular}
\label{array_simu_geometry}
\end{table}

In each case the total volume of the module
is kept {the} same. A gap of 5 mm was kept between the individual 
elements in all the simulations, though the choice of the gap size 
would depend upon the support structure of the individual elements. The 
fluence in each direction was 
kept proportional to the cosine of the angle between source direction 
and the local normal to the sphere surface. The radius of the sphere 
was kept much larger than the element size to ensure uniform illumination 
of the entire module. The detector multiplicity was defined as the number of elements in an event where the deposited energy is larger than the preset threshold of 10 
keV. 

Photon energies considered cover the range of interest for natural background radiations. The most 
prominent gamma radiations are from the $^{238}$U and $^{232}$Th series and 
$^{40}$K decay, with the maximum energy of 2615 keV from the $^{208}$Tl decay. Higher energy gamma-rays 
also exist, for example the 3183 keV from $^{214}$Bi, but 
the branching ratio is negligible (0.00133\%). {A} large number of events (10$^7$) were generated to 
minimize the statistical fluctuations in the simulated data. 

\begin{figure}[h]
\begin{center}
\includegraphics[width=0.8\linewidth,keepaspectratio]{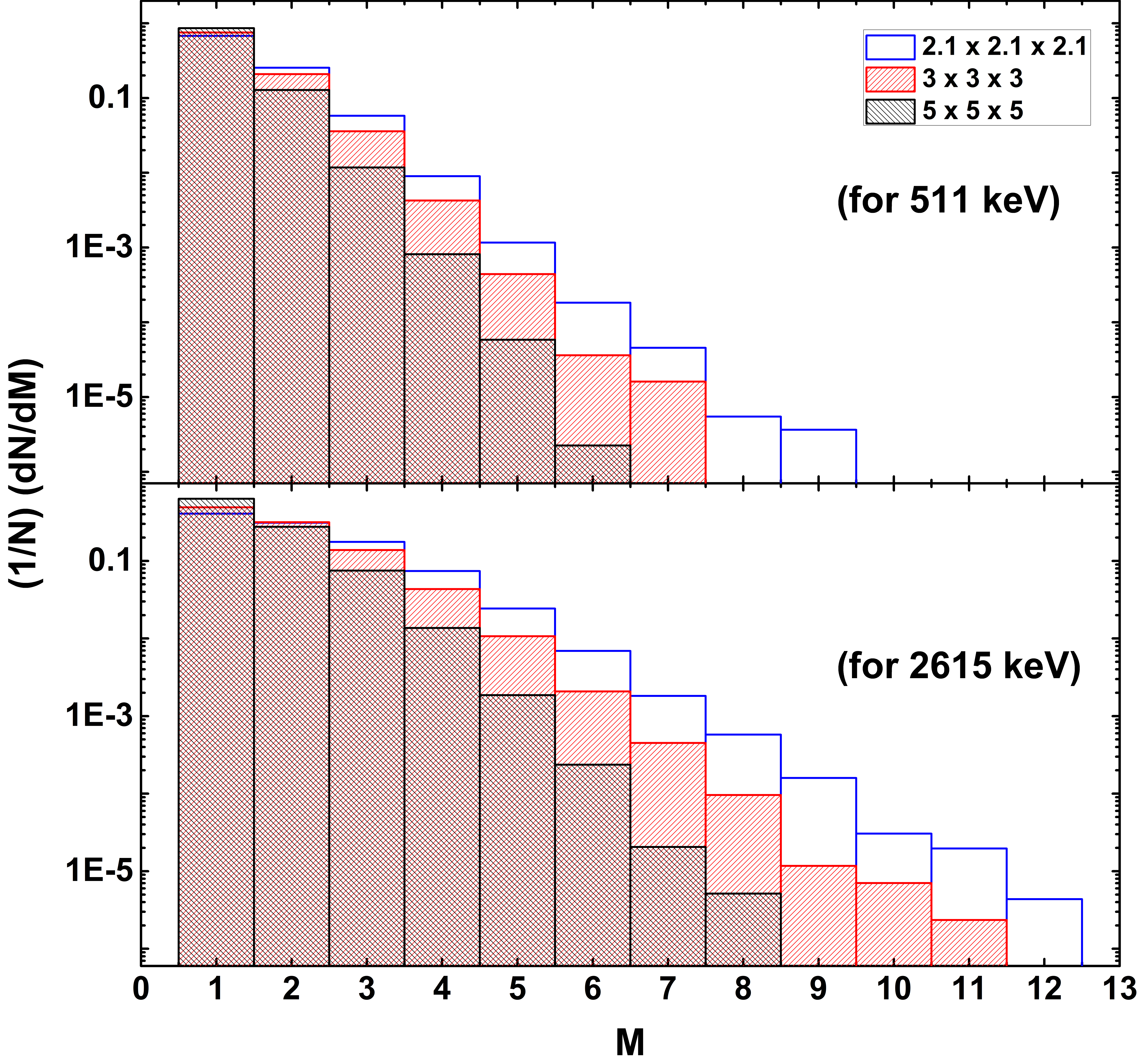}
\end{center}
\caption[Simulated multiplicity distribution for 511 keV and 2615
keV]{Simulated multiplicity distribution for $E_\gamma$= 511 keV and 
2615 keV for different element sizes.}       
\label{multiplicity}
\end{figure}
In each case the total volume of the module
is kept same. A gap of 5 mm was kept between the individual 
elements in all the simulations, though the choice of the gap size 
would depend upon the support structure of the individual elements. The 
fluence in each direction was 
kept proportional to the cosine of the angle between source direction 
and the local normal to the sphere surface. The radius of the sphere 
was kept much larger than the element size to ensure uniform illumination 
of the entire module. The detector multiplicity was defined as the number of elements in an event where the deposited energy is larger than the preset threshold of 10 
keV. 

The multiplicity distribution of the detector for 511 keV and 2615 keV is shown 
for different configurations in Figure~\ref{multiplicity}. The multiplicity 
distribution is narrower for larger element size and lower photon 
energy. Table~\ref{average_M} shows the average multiplicity and the percentage of photopeak events for $M=1$ for 
511 and 2615 keV for different element sizes. It can be clearly seen that the photopeak efficiency is higher for larger crystal size. However, the number of events that can be rejected on the basis of $M>1$ is higher for smaller crystal size. 
\begin{table}[h]
\caption[Average hit multiplicity ($<M>$) and
($\sigma_{<M>}$) and M=1 photopeak]{Average multiplicity ($<M>$) with its standard deviation 
($\sigma_{<M>}$) and the percentage of photopeak events ($M=1$), for 511 keV and 2615 keV for different 
element sizes.} 
\centering
\begin{tabular}{c c c c c}
\hline\hline
Energy & Element size & $<M>$ & $\sigma_{<M>}$ & $N^p_{M=1}/{N_{M=1}}$\\[0.5ex]
(keV) & (cm) & & & ($\%$) \\[0.5ex]
\hline\hline
511 & 2.1 & 1.4 & 2.3 & 76.1\\
 & 3 & 1.3 & 2.0 & 78.7\\
 & 5 & 1.2 & 2.0 & 81.2\\
\hline
2615 & 2.1 & 2.0 & 2.8 & 29.7\\
 & 3 & 1.8 & 2.6 & 37.2\\
 & 5 & 1.5 & 2.1 & 47.2\\
\hline \hline
\end{tabular}
\label{average_M}
\end{table}

\begin{figure}[ht]
\begin{center}
\includegraphics[scale = 0.4]{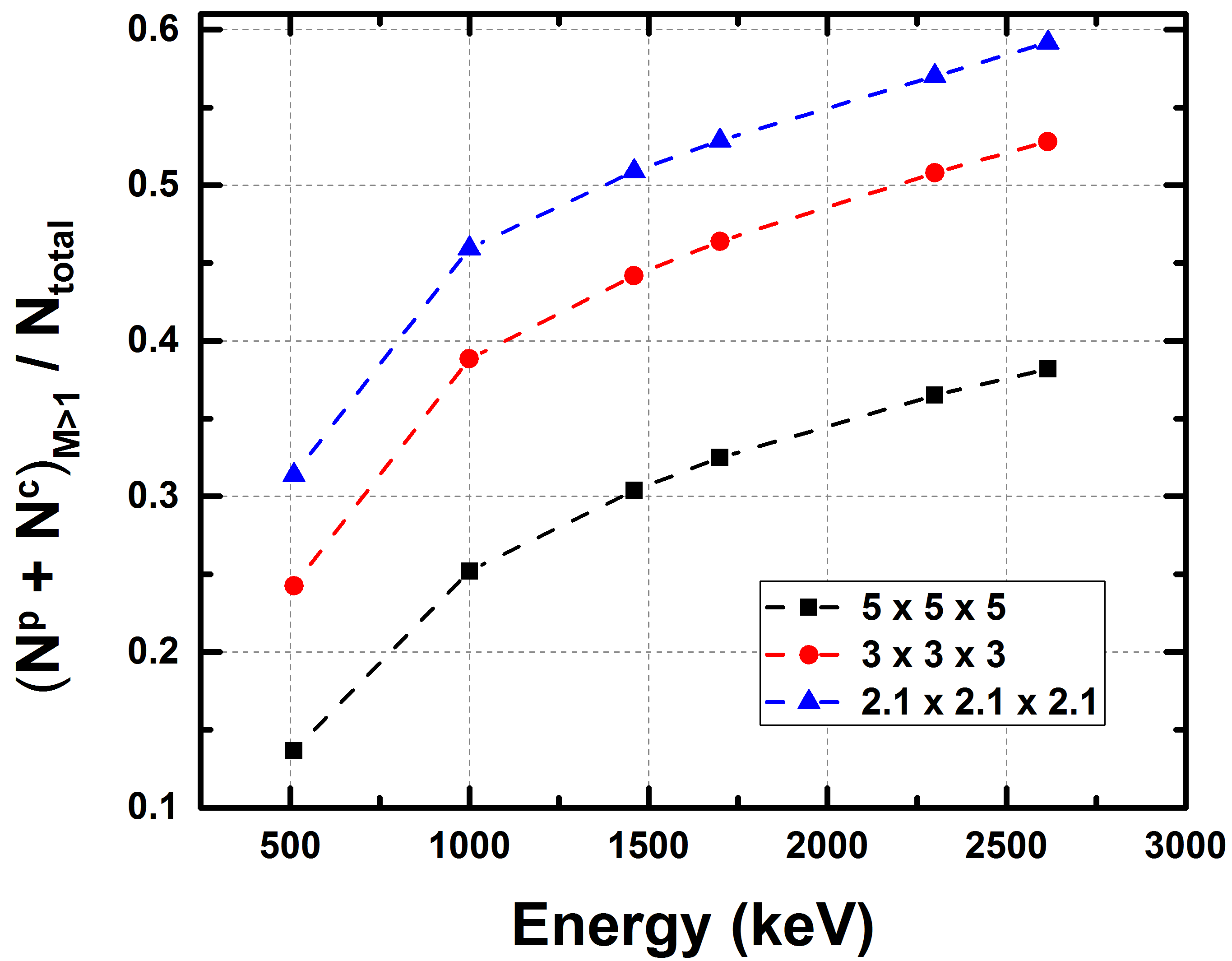}
\end{center}
\caption[The probability of $M >1 $ events as a 
function of energy]{The 
probability of $M >1 $ events as a 
function of photon energy for different element sizes (see text for 
details). Lines are only to guide the eye.}  
\label{multiplicity_greaterthan1_photons}
\end{figure}

Figure~\ref{multiplicity_greaterthan1_photons} shows the probability 
for discrimination of a photon based on a multiplicity condition of $M >1$. It 
is evident that smaller the size of an individual element, greater is 
the probability of discrimination of $M>1$ events. The 
background rejection ratio, defined as $N_{M>1}/N_{total}$, at 2615 keV 
is only $\sim$10\% worse for a=3 cm as compared to that for a= 2.1 cm, 
while it is about $\sim 35$\% worse for a=5 cm.

\begin{figure}[ht]
\begin{center}
\includegraphics[scale = 0.4]{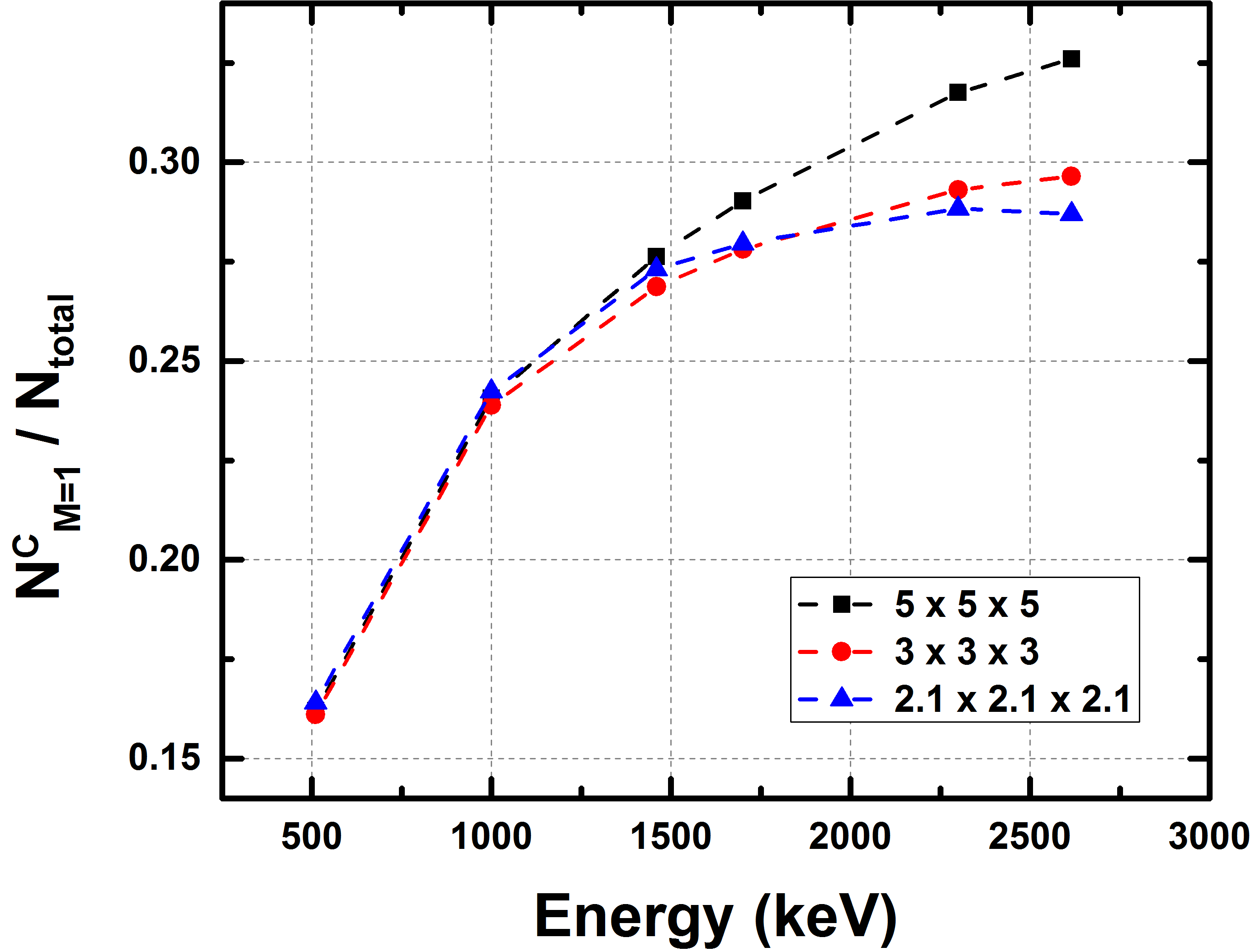}
\end{center}
\caption[The probability of $N^c_{M=1} $ events as a 
function of energy]{The 
probability of $N^c_{M=1}$ events as a 
function of photon energy for different element sizes (see text for 
details). Lines are only to guide the eye.}  
\label{multiplicity_Nb_M=1}
\end{figure}

Figure~\ref{multiplicity_Nb_M=1} shows the fraction of Compton 
scattered events with $M=1$. 
It can be seen that detector elements with 
a= 2.1 cm and a= 3 cm show very similar behaviour, while a=5 
cm is worse by about 20\%. It can be seen that the $N^c_{M=1}$ for 
a= 2.1 cm is greater than that for a = 3 cm and 5 cm at lower energies 
whereas, at higher energies ($\geq$ 1500 keV) the $N^c_{M=1}$ for a= 
2.1 cm is smaller than that for a = 3 cm and 5 cm elements. This can be understood 
considering the half thickness ($d_{1/2}$) for the absorption of 
photons in Tin compared to the element size. For E$_\gamma \geq$ 1500 
keV, the $d_{1/2}$ for Tin is more than 
2 cm~\cite{TOI}. Therefore, at higher energies the probability for 
multiple hits ($M>1$) increases for smaller crystal sizes, which is 
reflected in Figure~\ref{multiplicity_Nb_M=1} as the reduction of 
$N^c_{M=1}$ events.
	
The effect of inter detector spacing was also studied by varying the 
gap from 2 mm to 10 mm. A larger gap 
between the detector will increase the probability of gamma-rays escaping 
after Compton scattering. This will increase the $N^c_{M=1}$ events, 
thereby increasing the background with increase in the inter detector spacing. 
Though a minimum inter detector spacing is preferable, the actual spacing will be determined by the support structures for the detector. 

{It should be mentioned that while a smaller sized element would result in an improvement in the signal to noise ratio ($\Delta T_{signal}/\Delta T_{baseline} $), it would also result in an increase in the number of readout channels.} For the same total 
mass of $\sim$25 kgs, a Tin detector module with a = 2.1 cm (343 elements) will 
require $\sim$ 3 times more number of sensor readouts than that for a = 3 cm (125 
elements). A detector array with very large granularity (like 343 
elements) would require an upscaling of wirings inside the cryostat, 
cold electronics and data acquisition electronics at 
room temperature. Also, an increased number of sensors would correspond 
to more surrounding material (connecting wires etc.) which will 
contribute to the background, thereby adversely affecting the sensitivity.   
From the Figure~\ref{multiplicity_greaterthan1_photons} it can be deduced that 
the multiplicity discrimination for a = 2.1 cm at 2615 keV is only 
$\sim$10\% better than that for a = 3 cm. Thus, it appears that 3 x 3 
x 3 cm$^3$ element size provides the optimal granularity for the background 
discrimination of the gamma events and the number of readout channels. It is envisaged that the prototype of the TIN.TIN detector will consist 
of elements of sizes 3x3x3 cm$^3$ stacked in modules which will be arranged in a tower geometry. This will also facilitate upscaling for large mass detector. {It should be pointed out that only a minimal surface area closed geometry packing (cubical shape) has been considered for this work. However, it is possible that other detector geometries (e.g. rectangular cross-sections) may provide a better multiplicity discrimination for gamma events. Further studies on geometry optimization can be carried out, if background from surface events is precisely known.}



\section{Summary}

Monte Carlo simulations have been carried out to optimize the detector element size 
for photon background reduction based on hit multiplicity. The 
present studies indicate that a 3x3x3 cm$^3$ element for a detector module 
would be a suitable choice for calorimetry, background discrimination of 
gamma events and the number of readout channels. The suggested module design 
is a cubic array of 27 elements arranged in a 3x3x3 geometry. The gamma background in the region of interest ($>$2 MeV) can be reduced by $\sim$50\% by using the multiplicity information (M$\geq$1) from the segmented array of the optimized module.

\bibliographystyle{pramana}
\bibliography{references}

\begin{thebibliography}{99}
\bibitem{SuperK_Atmospheric_sy}
R.Wendell {\it et al.}, {\it Phys. Rev. D}, {\bf 81} 092004 (2010)
\bibitem{SNO_Solar_sy}
B. Aharmim {\it et al.}, {\it Phys.Rev. C}, {\bf 88} 025501 (2013) 
\bibitem{KamLAND_Reactor_sy}
A. Gando {\it et al.}, {\it Phys. Rev. D}, {\bf 83} 052002 (2011)
\bibitem{K2K_accelerator_sy}
M.H. Ahn {\it et al.}, {\it Phys. Rev. Lett.}, {\bf 90} 041801 (2003)
\bibitem{MINOS_accelerator_sy}
P. Adamson {\it et al.}, {\it Phys. Rev. Lett.}, {\bf 106} 181801 (2011)
\bibitem{LeptoGenesis_sy}
M. Fukugita and T. Yanagida, {\it Phys. Lett. B}, {\bf 174} 45 (1986)
\bibitem{RMPReview_2008}
F.T. Avignone, III, S.R. Elliott, and J. Engel, {\it Rev. Mod. Phys.}, {\bf 80} 481 (2008)
\bibitem{RPPTheory_2012}
J.D. Vergados, H. Ejiri, and F. Simkovic, {\it Rep. Prog. Phys.}, {\bf 75} 106301 (2012)
\bibitem{AHEPReview_2014}
O. Cremonesi and M. Pavan, {\it Advances in High Energy Physics}, {\bf 2014} 951432 (2014)
\bibitem{Naba_Pramana}
N.K. Mondal, {\it Pramana - Journal of Physics}, {\bf 79} 1003 (2012) 
\bibitem{Nanal_INPC13}
V. Nanal, {\it EPJ Web of Conferences}, {\bf 66} 08005 (2014)
\bibitem{Pavan}
M. Pavan {\it et al.}, {\it Eur. Phys. J A} {\bf 36} 159 (2008)
\bibitem{GEANT4}
S. Agostinelli {\it et al.}, {\it Nucl. Instrum. Methods Phys. Res. Sect. A} {\bf 506} 250 (2003)
\bibitem{TOI}
R.B. Firestone and V.S. Shirley, {\it Table Of Isotopes}, J. Wiley \& Sons, 8th Ed (1999) 

\end{thebibliography}

\end{document}